# RADIX-2 FAST HARTLEY TRANSFORM REVISITED

de Oliveira, H.M.; Viviane L. Sousa, Silva, Helfarne A.N. and Campello de Souza, R.M.

*Federal University of Pernambuco*
Departamento de Eletrônica e Sistemas - Recife - PE, Brazil
e-mail: hmo@ufpe.br ricardo@ufpe.br

**ABSTRACT**- *A Fast algorithm for the Discrete Hartley Transform (DHT) is presented, which resembles radix-2 fast Fourier Transform (FFT). Although fast DHTs are already known, this new approach bring some light about the deep relationship between fast DHT algorithms and a multiplication-free fast algorithm for the Hadamard Transform.*

## 1. INTRODUCTION

Discrete transforms provide a key tool in Signal Processing. A striking example is the Discrete Fourier Transform (DFT). Another relevant example concerns the Discrete Hartley Transform (DHT) [1], the discrete version of the integral transform introduced by R.V.L. Hartley in [2]. Besides its numerical side appropriateness, the DHT has proven over the years to be a powerful tool [3-5]. It was originally seen having connection to the physical world only via Fourier transform, but the DHT have found many interesting applications. The successful use of transform techniques relies on the existence of the so-called fast transforms [6]. Fast Hartley transforms also do exist and are deeply connected to the DHT applications [7,8]. Recent promising outcome of discrete transforms concern the use of finite field Hartley transforms [9] to design digital multiplexes, efficient multiple access systems [10] and multilevel spread spectrum sequences [11]. Very efficient algorithms such as Prime Factor Algorithm (PFA) or Winograd Fourier Transform Algorithm (WFTA) have also been used [13,14]. The discrete Hartley transform of a signal $v_i$, $i=0,1,2,...,N$-1 is defined as

$$V_k := \sum_{i=0}^{N-1} v_i \cdot cas\left(\frac{2\pi ki}{N}\right), \quad k=0,1,2,...,N\text{-}1, \qquad (1)$$

where $cas(x)=cos(x)+sin(x)$ is the "cosine and sine" Hartley symmetric kernel. There is a simple relationship between the DHT and the DFT of a given real discrete signal $f_i$, $i=0,1,...,N$-1:

If $f_i \leftrightarrow F_k$ is a DFT pair and $f_i \leftrightarrow H_k$ is the corresponding DHT pair, then [3] $\forall k$ $H_k = \Re e F_k - \Im m F_k$ and

$$F_k = \frac{1}{2}\left[(H_k + H_{N-k}) - j(H_k - H_{N-k})\right]$$

Therefore, a fast algorithm for the DHT is also a FFT for the DFT and vice versa (see Corollary 6.9 [15]). Besides being a real transform, the DHT is also *involutionary*, i.e.; the kernel of the inverse transform is exactly the same as that one of the direct transform (self-inverse transform). A radix-2 algorithm for DHT was introduced in [16]. Recently new fast algorithms have been introduced which are based on Hadamard decomposition [16]. There is some relationship between discrete Hartley and Hadamard transforms. In order to investigate such a link, a fast Hadamard transform is presented and a new derivation of the radix-2 Hartley transform is made. The revisited radix-2 clarifies the connection by showing that the "skeleton" of both algorithms is essentially the same.

## 2. A FAST HADAMARD TRANSFORM ALGORITHM

The attractiveness of Hadamard Discrete Transforms (HT) is essentially due to its low complexity when compared to other discrete transforms. The present paper investigates architecture for the HT that is suitable for implementations based on Digital Signal Processors (DSPs) or low-cost dedicated integrated circuits. <u>Hadamard Decomposition</u> is a procedure where the HT calculations are performed by means of decomposition into standard 2-HT sub-blocks (HT of length 2). This fast algorithm exploit successive breaking of the input signal into two components, reducing the transform blocklength by half in each step. Figure 1 illustrates the decomposition process for an 8-HT.

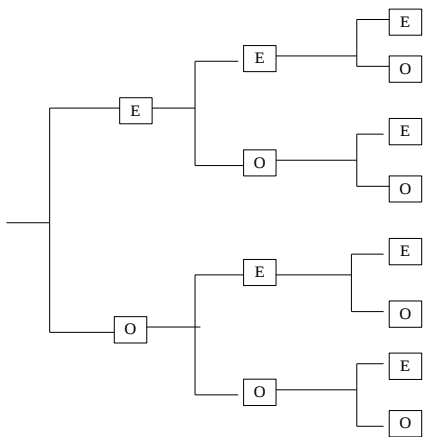

**Figure 1. Structure of the Hadamard tree decomposition by slitting blocks into even and odd components.**

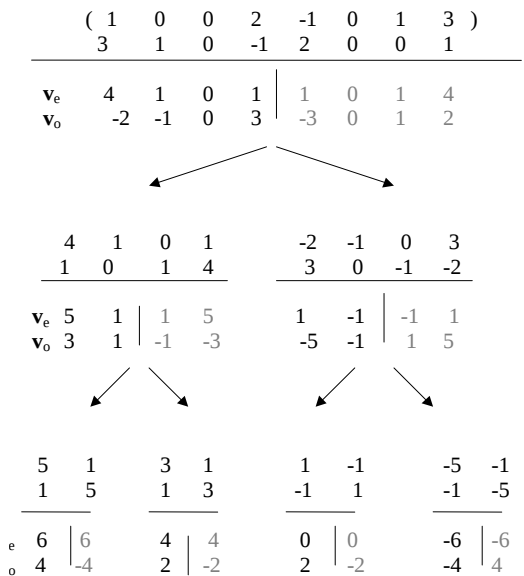

**Figure 2. A naive example of HT decomposition for *N*=8.**

Schemes based on cells somewhat similar to classical Cooley-Tukey butterflies are presented for several transform blocklengths. The basic structure of such an algorithm seems to be some kind of "skeleton" in assembling other discrete transforms such as the DFT and the DHT.

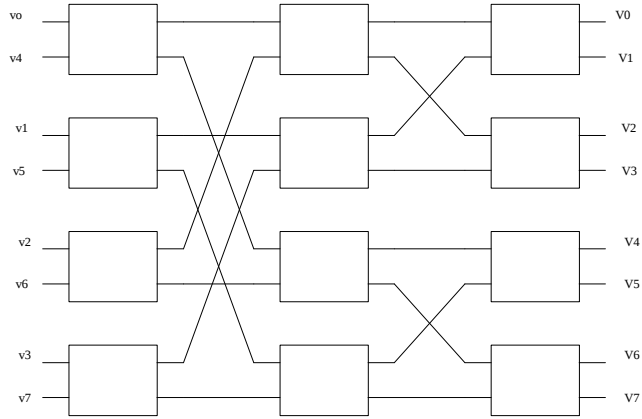

**Figure 3. A HT scheme with cells similar to "butterflies" of classical FFT by Cooley-Tukey**

In this approach, nested algorithms have been obtained for computing HT, which allow a great choice for the transform blocklength, by using the *same hardware*. This opens a window on constructing nested algorithms for other discrete transforms.

## 3. A RADIX-2 FHT ALGORITHM

Let us consider Discrete Hartley Transforms of length $N=2^m$, $m$ integer. Given a signal $v=\{v_0,v_1,...,v_{N-1}\}$, the DHT spectrum of **v** is the vector $V=(V_0,V_1,V_2,...,V_{N-1})$. We first break the input signal in two parts: $v=\{v_0, v_1,..., v_{N/2-1} \mid v_{N/2}, v_{N/2+1}..., v_{N-1}\}$. In order to find out a FHT algorithm we begin with the definition of "odd" and "even" half-wave signals:

**Definition 1.** *Given a discrete signal **v** with blocklength N, the even-half-wave-part of **v** is defined to be a vector $v_e=(v_{e0}, v_{e1}, ..., v_{eN-1})$ according to*

$$v_{ei} = \begin{cases} v_i + v_{N/2+1} & 0 \leq i \leq \frac{N}{2}-1 \\ v_{e(i \bmod N/2)} & \frac{N}{2} \leq i \leq N-1 \end{cases},$$

*and an odd-half-wave-part defined to be a vector $v_o=(v_{o0}, v_{o1}, ..., v_{oN-1})$ according to*

$$v_{oi} = \begin{cases} v_i - v_{N/2+1} & 0 \leq i \leq \frac{N}{2}-1 \\ -v_{o(i \bmod N/2)} & \frac{N}{2} \leq i \leq N-1 \end{cases}. \blacksquare$$

Clearly $\mathbf{v} = \frac{1}{2}(\mathbf{v_e} + \mathbf{v_o})$, that is, $v_i = \frac{1}{2}(v_{ei} + v_{oi})$.

Following this decomposition the Hartley spectrum (Eqn. 1) is given by

$$2V_k = \sum_{i=0}^{N-1}(v_{ei} + v_{oi})cas\left(\frac{2\pi ki}{N}\right). \tag{2}$$

Carrying out the summation in two parts:

$$2V_k = \sum_{i=0}^{\frac{N}{2}-1}(v_{ei}+v_{oi})cas\left(\frac{2\pi ki}{N}\right) + \sum_{i=\frac{N}{2}}^{N-1}(v_{ei}+v_{oi})cas\left(\frac{2\pi ki}{N}\right).$$

Changing the index of the second summation $i'=i-N/2$, it follows that

$$2V_k = \sum_{i=0}^{\frac{N}{2}-1}(v_{ei}+v_{oi})cas\left(\frac{2\pi ki}{N}\right) + \sum_{i'=0}^{\frac{N}{2}-1}(v_{e_{i'+N/2}}+v_{o_{i'+N/2}})cas\left(\frac{2\pi ki'}{N}+\pi k\right).$$

But $cas\left(\frac{2\pi ki'}{N}+\pi k\right) = (-1)^k cas\left(\frac{2\pi ki'}{N}\right)$ and it follows from Definition 1 that $v_{e_{i'+N/2}} = v_{e_{i'}}$ and $v_{o_{i'+N/2}} = -v_{o_{i'}}$. Therefore

$$2V_k = \sum_{i=0}^{\frac{N}{2}-1}(v_{ei}+v_{oi})cas\left(\frac{2\pi ki}{N}\right) + (-1)^k \sum_{i'=0}^{\frac{N}{2}-1}(v_{e_{i'}}-v_{o_{i'}})cas\left(\frac{2\pi ki'}{N}\right). \quad (3)$$

Now we consider half the vectors $v_e$ and $v_o$:

**Definition 2**. *Let $u_e=(u_{e0}, u_{e1},..., u_{eN/2-1})$ and $u_o=(u_{o0}, u_{o1},...,u_{oN-1})$ with components $u_{ei}=v_{ei}$ and $u_{oi}=v_{oi}$, $i=0,1,...,N/2-1$.* ■

The Discrete Hartley Transform of such vectors are $\{u_e\} \leftrightarrow \{U_e\}$ and $\{u_o\} \leftrightarrow \{U_o\}$, i.e.,

$$U_{ek} = \sum_{i=0}^{\frac{N}{2}-1} v_{ei} cas\left(\frac{2\pi ki}{N/2}\right) \text{ and } U_{ok} = \sum_{i=0}^{\frac{N}{2}-1} v_{oi} cas\left(\frac{2\pi ki}{N/2}\right).$$

Equation (3) becomes

$$V_k = U_{ek/2}\left\lceil\frac{1+(-1)^k}{2}\right\rceil + U_{ok/2}\left\lceil\frac{1-(-1)^k}{2}\right\rceil. \quad (4)$$

We have then proved the following proposition:

**Proposition 1.** *The computing of an N-blocklength Discrete Hartley Transform of a vector $v$, can be done in term of DHTs of length N/2 according to:*

(i) The Hartley spectral components of pair indexes, $V_{2k}$, of the vector v is $V_{2k} = U_{e_k}$,

(ii) The Hartley spectral components of odd indexes, $V_{2k+1}$, of the vector $v$ is $V_{2k+1} = U_{o_{k+\frac{1}{2}}}$ ■

The pair harmonic components are immediately obtained but odd harmonics require further calculations. Applying the DHT property $\cos\left(\frac{2\pi k_0 i}{N}\right)g_i + \sin\left(\frac{2\pi k_0 i}{N}\right)g_{-i} \leftrightarrow G_{k-k_0}$ for the signal $U_{o_{k+\frac{1}{2}}}$,

$$\cos\left(\frac{\pi i}{N}\right)u_{oi} - \sin\left(\frac{\pi i}{N}\right)u_{o-i} \leftrightarrow U_{o_{k+\frac{1}{2}}} = V_{2k+1}. \quad (5)$$

Therefore, the odd harmonic components are calculated from $u_O$ after some adjustment terms. Evaluating the multiplicative complexity. The above proceed can iterative be applied since that $N$ is a power of two. We arrive then to the block diagram shown in Figure 4 below.

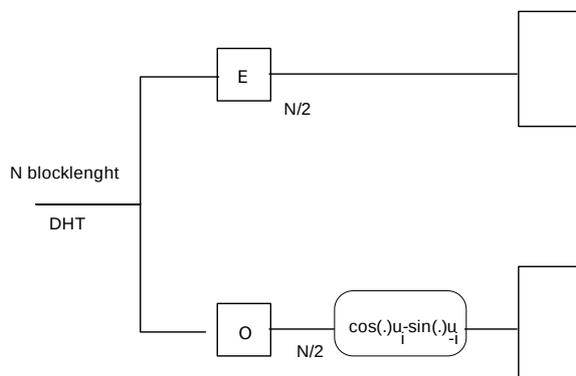

**Figure 4. A step of the Fast Discrete Hartley Transform based on even-odd decomposition.**

Since that each step reduces the block length to a half, there is $\log_2 N$ steps. Those of the adjustments, i.e., $2N$ multiplications give the number of multiplications in each step. Therefore, we need $2N\log_2 N$ multiplications the same as the classical Cooley-Tukey FFT.

## 4. CONCLUDING REMARKS

A decisive factor for applications of the DHT has been the existence of the so-called fast transforms for computing it. The algorithm described in this paper is equivalent to a frequency-decimation radix-2 fast algorithm. It is based on a decomposition of the time-signal into even and odd components. Hence, it enables a better understanding on the deep relationship between Hartley Transform and Hadamard Transform.